# The Didactical Relevance of the Pauli Pascal Triangle

# Die didaktische Relevanz des Pauli-Pascal-Dreiecks[1]

Martin Erik Horn[2]

**Abstract**

In high school physics and physics at university level anti-commutative and non-commutative quantities play an outstanding role at the theoretical description of physical relations. But while using commutative quantities, the basic formal relations of mathematical physics are introduced clearly and step-by-step, a similar clear presentation of non-commuatative relations with didactical sensible examples is lacking.

On the basis of a q-analog of the Pascal triangle (sometimes called quantum Pascal triangle) this paper discusses whether a less abstract introduction into non-commutative relations is possible. The didactical relevance of this approach is analysed. In an attachment all three triangles of the Pauli Pascal plane are presented.

**Kurzfassung**

In der Hochschulphysik spielen anti-kommutative und nicht-kommutative Größen eine herausragende Rolle bei der theoretischen Formulierung physikalischer Zusammenhänge. Doch während bei Betrachtung kommutativer Größen die der mathematischen Physik zugrundeliegenden Beziehungen anschaulich aufbauend eingeführt werden, fehlt eine ähnlich anschauliche und an Anwendungsbeispielen orientierte Einführung nicht-kommutativer Beziehungen.

Anhand des q-Analogons des Pascal Dreiecks (auch Quanten-Pascal-Dreieck genannt) wird diskutiert, ob eine weniger abstrakte Hinführung gelingen kann, und welche didaktische Relevanz diesem Ansatz zukommt. Im Anhang werden die drei Dreiecke der vollständigen Pauli-Pascal-Ebene vorgestellt.

**Contents**

**Inhalt**

[1] The German version of this paper will be published (without attachment) in:
Dietmar Höttecke (Ed.): Tagungsband der GDCP-Jahrestagung 2006 in Bern, Beitrag P 06.

[2] Address: FHW Berlin, Badensche Str. 50/51, Fach # 112, 10825 Berlin, Email: horn@fhw-berlin.de,
University of Potsdam, Am Neuen Palais 10, 14469 Potsdam, Germany, Email: marhorn@uni-potsdam.de.

## 1. Historical Background

At the development of modern quantum mechanics Dirac [4] formulated the demand for an original and actual quantum algebra. This quantum algebra should be built on non-commutative q-numbers instead of classical c-numbers. But such a mathematical language – the mathematics of q-functions – had already been developed by Rothe, Heine, Cauchy and others over hundred years before [2, p. 10].

It seems psychologically interesting that even the designation with the letter q, which should be interpreted later as prefix "quantum", had been used long before the development of quantum mechanics [3]. The mathematics of quantum mechanics is much older than the physical construct of quantum mechanics.

## 2. Non-commutative relations in physics

Quantum mechanics may be an important example for the application of mathematical relations which are not commutative but quantum mechanics is by far not the only one. Indeed already before publishing his theory of extensions Grassmann describes the theory of the tides on the basis of an outer multiplication of vectors in his thesis at the examination for secondary school teachers of physics [8].

Hestenes also uses an associative, but not commutative geometric algebra to describe classical as well as quantum mechanical physical phenomena mathematically [6]. The non-commutativity of basic entities in geometric algebra is one of the reasons for the effective and structurally clear formulation. And it is a theoretical formulation in the sense of a mathematical language of physics.

Therefore non-commutative elements and non-commutative relations play an important role especially in classical physics at theoretically formulating and describing physical phenomena.

For example it makes a strong difference, whether one turns to right first and then turns to the left, which results in a counterclockwise movement, or whether one first turns to the left and only after that then turns to the right, which results in a clockwise movement. Consequently directions (or orthogonal vectors) anticommute. And even though it may be possible to model directions mathematically by using commutative quantities, this doesn't seem appropriate to the anti-commutative structure of the phenomenon.

## 3. Introduction of commutative and non-commutative relations

In primary and secondary school as well as in high school or university physics commutative quantities and the relations between them are presented in the

## 1. Historischer Hintergrund

Während der Genese der modernen Quantenmechanik formuliert Dirac [4] die Forderung nach einer eigenständigen Quantenalgebra mit nichtkommutativen q-Zahlen anstelle der klassischen c-Zahlen. Doch eine solche mathematische Sprache, die Mathematik der q-Funktionen, wurde bereits über 100 Jahre zuvor von Rothe, Heine, Cauchy und anderen entwickelt Andrews [2, S. 10].

Psychologisch interessant scheint, dass selbst die Bezeichnung mit dem Buchstaben q, der später als Vorsilbe „quanten..." interpretiert werden sollte, bereits weit vor Entwicklung der Quantenmechanik genutzt wird [3]. Die Mathematik der Quantenmechanik ist weit älter als das physikalische Konstrukt Quantenmechanik.

## 2. Nichtkommutativitäten in der Physik

Die Quantenmechanik ist zwar ein wichtiges Beispiel für die Anwendung nichtkommutativer mathematischer Beziehungen in der Physik, bei weitem jedoch nicht das einzige. So beschreibt Grassmann noch vor Veröffentlichung seiner Ausdehnungslehre in einer Prüfungsarbeit zur Erlangung der Lehrbefähigung für den Physikunterricht die Theorie von Ebbe und Flut auf der Grundlage der äußeren Multiplikation von Vektoren [8].

Auch Hestenes nutzt die nichtkommutative, jedoch assoziative Geometrische Algebra zur Beschreibung klassischer wie auch quantenmechanischer physikalischer Phänomene [6]. Die Nichtkommutativität der Basisgrößen in der Geometrischen Algebra ist einer der Gründe für die effektive und strukturell klare theoretische Beschreibung im Sinne einer mathematischen Sprache der Physik.

Nichtkommutative Größen und nichtkommutative Beziehungen spielen bei der theoretischen Aufarbeitung und Beschreibung physikalischer Phänomene deshalb gerade auch im Bereich der klassischen Physik eine wichtige Rolle.

Es macht beispielsweise einen deutlichen Unterschied, ob man zuerst rechts abbiegt und sodann links, was eine Bewegung im Gegen-Uhrzeigersinn zur Folge hat, oder ob man zuerst links abbiegt und sodann rechts mit der Folge einer Bewegung im Uhrzeigersinn. Die Richtungen vertauschen somit antikommutativ, und obwohl es möglich ist, diese mit kommutativen Größen mathematisch zu modellieren, erscheint es aufgrund der antikommutativen Struktur des Phänomens nicht unbedingt sinnvoll.

## 3. Einführungsweise in kommutative und nichtkommutative Beziehungen

In Grundschule, Sekundarstufe I, Sekundarstufe II und im Grundstudium Physik werden kommutative Größen und deren Beziehungen sehr, sehr ausführ-lessons in great detail, almost constantly, and often only in very small steps and with many practical exercises. Thus to learn the basic arithmetical operations at school already means at the same time to learn commutative relations. This is didactically of course sensible.

Compared with that the teaching of some few examples of non-commutativity is not given much attention in school and high school and only a small amount of attention in basic courses of university in relative terms. Content-oriented there is a clear difference in the level of teaching and explaining commuting and non-commuting relations too.

While commutative quantities are introduced at a very basic and elementary level, the presentation of non-commutative quantities mostly has abstract, cognitively demanding and very often not illustrative standards. Even introductory texts of q-deformed mathematics – that is, the mathematics of quantum mechanical relations – show no detailed explanations, but abstract and very dense depictions using hypergeometric functions (see for example [2], [5] or [7]).

Commuting and non-commuting quantities are introduced at different times. This may be one reason for the difference of the teaching level. Another reason surely is that different learning groups are affected. Yet the conclusion remains: There is a lack of a concept to introduce non-commutative relations for students who are not mathematics students. Such a concept should be similar to concepts of teaching basic arithmetical operations.

## 4. The Pascal triangle

The different levels of introducing commutative and non-commutative mathematics are visible at numerous mathematical structures. For example the Pascal triangle plays a key role in commutative mathematics as cognitive attractor. It is one of the oldest mathematical objects, which was described lich, nahezu permanent, oft sehr kleinschrittig und unter Einbezug zahlreicher Übungsphasen im Unterricht behandelt. So ist schon das Erlernen der Grundrechenarten gleichzeitig ein Erlernen kommutativer Beziehungen. Das ist aus didaktischer Sicher nur zu begrüßen.

Dagegen nimmt die Vermittlung einiger weniger Beispiele von nichtkommutativen Beziehungen einen sehr geringen Raum in der Sekundarstufe II und nur einen relativ geringen Raum im Grundstudium ein. Es besteht auch inhaltlich ein deutlicher Niveauunterschied bei der Vermittlung kommutativer und nichtkommutativer Zusammenhänge.

Während kommutative Größen sehr elementar eingeführt werden, erfolgt die Behandlung nichtkommutativer Größen meist auf einem abstrakten, kognitiv anspruchsvollen und oft unanschaulichen Niveau. Selbst einführende Texte in die q-deformierte Mathematik, also die Mathematik quantenmechanischer Beziehungen, (z. B. [2], [5] oder [7]) verfolgen kein ausführlich erklärendes Vorgehen, sondern bieten abstrakte und sehr kompakte Darstellungen unter Nutzung von hypergeometrischen Funktionen.

Gründe für diese Diskrepanz mögen in unterschiedlichen Einstiegszeitpunkten und unterschiedlichen Zielgruppen zu finden sein. Das Fazit aber bleibt: Es fehlt ein dem Erlernen und Einüben der Grundrechenarten vergleichbarer Ansatz, um Nichtmathematiker wie beispielsweise Physikstudentinnen und -studenten an nichtkommutative Beziehungen heranzuführen.

## 4. Das Pascal-Dreieck

Die unterschiedlichen Einstiegsniveaus in die kommutative und nichtkommutative Mathematik werden an zahlreichen mathematischen Strukturen sichtbar. So kommt beispielsweise dem Pascal-Dreieck in der kommutativen Mathematik eine Schlüsselrolle als kognitiver Attraktor zu. Es ist eines der ältesten ma-and explored systematically. And it was already known long before Pascal in Chine and India.

| | |
|---|---|
| 1 | $(a + b)^0 = 1$ |
| 1 1 | $(a + b)^1 = 1a + 1b$ |
| 1 2 1 | $(a + b)^2 = 1a^2 + 2ab + 1b^2$ |
| 1 3 3 1 | $(a + b)^3 = 1a^3 + 3a^2b + 3ab^2 + 1b^3$ |
| 1 4 6 4 1 | $(a + b)^4 = 1a^4 + 4a^3b + 6a^2b^2 + 4ab^3 + 1b^4$ |
| 1 5 10 10 5 1 | |
| 1 6 15 20 15 6 1 | etc… |
| 1 7 21 35 35 21 7 1 | |

***Fig. 1:*** *Construction of the Pascal triangle using binomial relations.*
***Abb. 1:*** *Konstruktion des Pascal-Dreiecks mit Hilfe binomischer Beziehungen.*

The Pascal triangle codifies the binomial coefficients and consequently the binomial formulae. In addition it is structurally deeply linked to the Taylor expansion. And it offers a didactical introduction into hypergeometric functions which is easy to implement and can be taught in a straightforward way.

This should not be underestimated because high school and university physics is directly influenced by that. A didactical analysis shows that according to Sawyer [9]: „in fact there must be many universities today where 95 per cent, if not 100 per cent, of the functions studied by physicists, engineering and even mathematics students are covered by this single symbol ${}_2F_1[a, b; c; x]$." Therefore hypergeometric functions can be of direct use as structural frame of mathematical physics [10].

## 5. The Pauli Pascal triangle

To make the introduction of the theoretical description of quantum mechanics conceptually easier, a quantum Pascal triangle [3] can be constructed. Following Baez this q-Pascal triangle shows "Pascal's triangle in a magnetic field". The parameter $\mathbf{q}$ refers to the relation between the quantities $\mathbf{x}$ and $\mathbf{y}$. They anticommute at an interchange of position and therefore satisfy $\mathbf{xy = q\ yx}$ . Using the q-Pascal triangle the teaching and learning of non-commutative relations can be as illustrative and cognitive attractive as in the case of commutative relations.

This is especially easy to implement for the case of $\mathbf{q = -1}$ . The anticommuting basic elements of geometric algebra can be seen as such q-deformed quantities. For example the Pauli matrices, which can form a basis of the three-dimensional Euclidean vector space, show the following multiplication rule: $\sigma_x \sigma_y = -\sigma_y \sigma_x$ [6]. The Pauli Pascal triangle therefore structurally codifies binomial coefficients which are q-deformed with $\mathbf{q = -1}$ . It has a threefold structure (see figure 2). This structure can be used as a didactical route to the q-binomial theorem. And it provides a didactical bridge from physical phenomena which can be described with commuting quantities to phenomena which are modelled with anticommuting quantities.

Equally interesting should be, that the basic mathematical relations can be found without the use of differential equations. This is just the level, which students of introductory courses looking for a clear and graphic framework need. The Pauli Pascal triangle paves the road to the more general quantum Pascal triangle. It again provides a didactical door opener for hypergeometric functions, which play a central role in physical modelling processes already described exemplarily by Andrews [6].

thematisch untersuchten und beschriebenen Objekte und war schon lange vor Pascal in China und Indien bekannt.

Mathematisch kodifiziert das Pascal-Dreieck die Binomialkoeffizienten und infolge dessen die Binomischen Formeln. Darüber hinaus ist es eng mit der Taylor-Entwicklung verknüpft und bietet einen didaktisch leicht zugänglichen Einstieg in die hypergeometrischen Funktionen.

Dies ist aus Sicht einer Hochschul-Physikdidaktik nicht zu unterschätzen, denn so Sawyer [9]: „Tatsächlich muss es heute zahlreiche Universitäten geben, in denen 95 %, wenn nicht gar 100 % der Funktionen, die von Physik-, Ingenieurs- oder sogar Mathematikstudenten während ihres Studiums behandelt werden, durch dieses eine Symbol abgedeckt werden: ${}_2F_1[a, b; c; x]$." Deshalb können hypergeometrische Funktionen als strukturierendes Gerüst der mathematischen Physik [10] dienen.

## 5. Das Pauli-Pascal-Dreieck

Um einen Einstieg in die theoretische Beschreibung der Quantenmechanik zu erleichtern, kann ein Quanten-Pascal-Dreieck [3] konstruiert werden. Dieses q-Pascal-Dreieck beschreibt nach Baez „Pascals Dreieck in einem magnetischen Feld". Der Parameter q bezieht sich dabei auf Größen $\mathbf{x}$ und $\mathbf{y}$, die der Vertauschungsrelation $\mathbf{xy = q\ yx}$ unterliegen. Mit Hilfe des q-Pascal-Dreiecks kann das Erlernen nicht-kommutativer Beziehungen kognitiv ansprechend und ähnlich anschaulich wie im kommutativen Fall gestaltet werden.

Insbesondere für den Fall $\mathbf{q = -1}$ ist dies leicht umzusetzen. Die antikommutierenden Basiseinheiten der Geometrischen Algebra können als solche q-deformierte Größen aufgefasst werden. Beispielsweise gilt für die Pauli-Matrizen als Basisvektoren des dreidimensionalen euklidischen Raumes $\sigma_x \sigma_y = -\sigma_y \sigma_x$ [6]. Das Pauli-Pascal-Dreieck kodifiziert deshalb die mit $\mathbf{q = -1}$ deformierten Binomialkoeffizienten und zeigt eine dreiwertige Struktur (siehe Abbildung 2). Diese Struktur kann als anschauliche Hinführung zum q-Binomialtheorem dienen und bietet eine Brücke von kommutativ beschreibbaren Phänomenen hin zur mathematischen Modellierung physikalischer Phänomene auf antikommutativer Grundlage.

Didaktisch interessant ist auch, dass die zugrunde liegenden Beziehungen ohne Differentialgleichungen und somit auf einem eingänglichen und anschaulichen Niveau erarbeitet werden können. Das Pauli-Pascal-Dreieck bietet einen Einstieg in das allgemeinere Quanten-Pascal-Dreieck. Dieses stellt wiederum einen Türöffner zu den q-deformierten hypergeometrischen Funktionen dar, deren Rolle im physikalischen Modellierungsprozess schon Andrews [1] exemplarisch herausgearbeitet hat.

1

1 1

1 0 1

1 1 1 1

1 0 2 0 1

1 1 2 2 1 1

1 0 3 0 3 0 1

1 1 3 3 3 3 1 1

1 0 4 0 6 0 4 0 1

1 1 4 4 6 6 4 4 1 1

0 5 0 10 0 10 0 5 0 1

1 5 5 10 10 10 10 5 5 1 1

$$(\sigma_x + \sigma_y)^0 = 1$$

$$(\sigma_x + \sigma_y)^1 = 1\sigma_x + 1\sigma_y$$

$$(\sigma_x + \sigma_y)^2 = 1\sigma_x^2 + 0\sigma_x\sigma_y + 1\sigma_y^2$$

$$(\sigma_x + \sigma_y)^3 = 1\sigma_x^3 + 1\sigma_x^2\sigma_y + 1\sigma_x\sigma_y^2 + 1\sigma_y^3$$

$$(\sigma_x + \sigma_y)^4 = 1\sigma_x^4 + 0\sigma_x^3\sigma_y + 2\sigma_x^2\sigma_y + 0\sigma_x\sigma_y^3 + 1\sigma_y^4$$

etc…

***Fig. 2:*** *Construction of the Pauli Pascal triangle using Pauli matrices.*
***Abb. 2:*** *Konstruktion des Pauli-Pascal-Dreiecks mit Hilfe der Pauli-Matrizen.*

## 6. Outlook: What can a didactical restructuring of high school physics achieve?

Theoretical models in mathematics and physics are free inventions of the human mind. Which of these invented mathematical and physical models are successful depend only on individual subjective and aesthetical preferences and on the practical relevance the scientific protagonists assign to the models in question. Just these are the questions didactical research among other things cares about and discusses critically.

Text and context of teaching and learning depend on one another. They are also deeply interrelated in connection with high school and university physics education. Lecturers in teaching methods and physics education researchers can contribute soundly to the question which model is of practical use and can be handled in a coherent modeling edifice.

Geometric algebra can be seen as a didactically reconstructed algebra of Pauli matrices. This shows that physics education researchers can give new impulses to rethink and reformulate physical relations. It is exactly our central didactical task and our didactical responsibility, also to reconstruct modern topics of physics *didactically*. And this means, to reconstruct theses topics *technically*, too.

## 6. Ausblick: Was kann eine Didaktik der Hochschulphysik leisten?

Theoretische Modellierungen in Mathematik und Physik sind freie Erfindungen des menschlichen Geistes. Welche mathematische Modellierung und welches physikalische Modell sich durchsetzt, entscheidet sich letztlich nur anhand subjektiver ästhetischer und praxisorientierter Prägungen der wissenschaftlichen Akteure. Genau das aber ist die Frage, mit der sich die Didaktik unter anderem auseinander setzt.

Text und Kontext des Lehrens und Lernens bedingen einander und stehen auch im Rahmen der Hochschul-Physikdidaktik in enger Beziehung. Welches Modell praktikabel und handhabbar ist, dazu können Physikdidaktikerinnen und -didaktiker plausible Beiträge liefern.

Die Geometrische Algebra als didaktisch aufbearbeitete Algebra der Pauli-Matrizen zeigt, dass Didaktiker bei der Aus- und Neuformulierung physikalischer Beziehungen Impulse setzen können. Es ist unser didaktisches Kerngeschäft, gerade auch moderne Themengebiete der Physik didaktisch zu rekonstruieren – und in Folge dessen die Physik (Text und Kontext bedingen sich) auch fachlich zu rekonstruieren.

## 7. Attachment

### I. q-numbers and q-binomial coefficients

The following q-deformed quantities can be defined as:

q-numbers:

## 7. Anhang

### I. q-Zahlen und q-Binomialkoeffizienten

q-deformierte Größen können folgendermaßen definiert werden:

q-Zahlen:

$$[n] = \frac{q^n - 1}{q - 1} = 1 + q + q^2 + \ldots + q^{n-1}$$

q-factorials:

q-Fakultäten:

$$[n]! = [1] \cdot [2] \cdot [3] \cdot \ldots \cdot [n]$$

q-binomial coefficients:

q-Binomialkoeffizienten:

$$\begin{bmatrix} n \\ m \end{bmatrix}_q = \frac{[n]!}{[m]!\,[n-m]!}$$

$$= \frac{(1+q)\cdot \ldots \cdot (1+q+q^2+\ldots+q^{n-1})}{(1+q)\cdot\ldots\cdot(1+q+q^2+\ldots+q^{m-1})\cdot(1+q)\cdot\ldots\cdot(1+q+q^2+\ldots+q^{n-m-1})}$$

$$= \frac{(1+q+q^2+\ldots+q^m)\cdot\ldots\cdot(1+q+q^2+\ldots+q^{n-1})}{(1+q)\cdot\ldots\cdot(1+q+q^2+\ldots+q^{n-m-1})}$$

With $\mathbf{q = -1}$ :

Für $\mathbf{q = -1}$ ergibt sich somit:

| | | |
|---|---|---|
| [1] = 1 | [1]! = [1] = 1 | |
| [2] = 0 | [2]! = [1][2] = 0 | |
| [3] = 1 | [3]! = [1][2][3] = 0 | |
| [4] = 0 | [4]! = [1][2][3][4] = 0 | etc... |

To prevent a division by zero when calculating binomial coefficients, problematic terms should be cancelled before inserting $\mathbf{q = -1}$ .
For example:

Um bei Berechnung der q-Binomialkoeffizienten eine Division durch Null zu vermeiden, sollte vor dem Einsetzen von $\mathbf{q = -1}$ gekürzt werden. Zum Bespiel berechnen sich:

$$\begin{bmatrix} 4 \\ 2 \end{bmatrix}_{-1} = \frac{(1+q)\cdot(1+q+q^2)\cdot(1+q+q^2+q^3)}{(1+q)\cdot(1+q)}$$

$$= \frac{(1+q+q^2)\cdot(1+q+q^2+q^3)}{(1+q)} = (1+q+q^2)\cdot(1+q^2) = (1-1+1)\cdot(1+1) = 2$$

$$\begin{bmatrix} 6 \\ 4 \end{bmatrix}_{-1} = \frac{(1+q)\cdot(1+q+q^2)\cdot(1+q+q^2+q^3)\cdot(1+q+q^2+q^3+q^4)\cdot(1+q+q^2+q^3+q^4+q^5)}{(1+q)\cdot(1+q)\cdot(1+q+q^2)\cdot(1+q+q^2+q^3)}$$

$$= \frac{(1+q+q^2+q^3+q^4)\cdot(1+q+q^2+q^3+q^4+q^5)}{(1+q)}$$

$$= (1+q+q^2+q^3+q^4)\cdot(1+q^2+q^4) = (1-1+1-1+1)\cdot(1+1+1) = 3$$

Mathematically this can be formulated as a limit:

Mathematisch ist dies einer Grenzwertbildung gleichwertig:

$$\begin{bmatrix} n \\ m \end{bmatrix}_{-1} = \lim_{q \to -1} \frac{(1+q+q^2+\ldots+q^m)\cdot\ldots\cdot(1+q+q^2+\ldots+q^{n-1})}{(1+q)\cdot\ldots\cdot(1+q+q^2+\ldots+q^{n-m-1})}$$

## II. The Pauli Pascal plane

Negative q-numbers are:

$$[-n]=\frac{q^{-n}-1}{q-1}=-q^{-n}\cdot\left(1+q+q^2+\ldots+q^{n-1}\right)=-\frac{1}{q}-\frac{1}{q^2}-\ldots-\frac{1}{q^n}=-q^{-n}[n]$$

This gives the binomial coefficients:

$$\begin{bmatrix}-n\\ m\end{bmatrix}_q=\begin{bmatrix}-n\\ -n-m\end{bmatrix}_q=\frac{[-n]!}{[m]!\,[-n-m]!}=\frac{[-n-m+1][-n-m+2]\cdot\ldots\cdot[-n]}{[m]!}$$

$$=(-1)^m\cdot q^{mn+\frac{m^2}{2}+\frac{m}{2}}\cdot\begin{bmatrix}n+m-1\\ m\end{bmatrix}$$

For example :

$$\begin{bmatrix}-4\\ 2\end{bmatrix}_{-1}=\begin{bmatrix}-4\\ -6\end{bmatrix}_{-1}=(-1)^2\cdot(-1)^9\cdot\begin{bmatrix}5\\ 2\end{bmatrix}_{-1}=-2$$

$$\begin{bmatrix}-6\\ 4\end{bmatrix}_{-1}=\begin{bmatrix}-6\\ -10\end{bmatrix}_{-1}=(-1)^4\cdot(-1)^{30}\cdot\begin{bmatrix}9\\ 4\end{bmatrix}_{-1}=6$$

## II. Die Pauli-Pascal-Ebene

Die negativen q-Zahlen sind:

Daraus ergeben sich die Binomialkoeffizienten:

Beispielsweise sind:

| | |
|---|---|
| 0 1 | 1 0 |
| –5 1 1 | 1 1 –5 |
| –4 0 1 | 1 0 –4 |
| –4 –4 1 1 | 1 1 –4 –4 |
| 0 –3 0 1 | 1 0 –3 0 |
| 6 –3 –3 1 1 | 1 1 –3 –3 6 |
| 3 0 –2 0 1 | 1 0 –2 0 3 |
| 3 3 –2 –2 1 1 | 1 1 –2 –2 3 3 |
| 0 1 0 –1 0 1 | 1 0 –1 0 1 0 |
| –1 1 1 –1 –1 1 1 | 1 1 –1 –1 1 1 –1 |

| |
|---|
| 1 |
| 1 1 |
| 1 0 1 |
| 1 1 1 1 |
| 1 0 2 0 1 |
| 1 1 2 2 1 1 |
| 1 0 3 0 3 0 1 |
| 1 1 3 3 3 3 1 1 |
| 1 0 4 0 6 0 4 0 1 |
| 1 1 4 4 6 6 4 4 1 1 |
| 1 0 5 0 10 0 10 0 5 0 1 |

***Fig. 3:*** *The three triangles of the Pauli Pascal plane.*
***Abb. 3:*** *Die drei Dreiecke der Pauli-Pascal-Ebene.*

These results indicate that a bilateral q-analog of the binomial theorem and Ramanujans ${}_1\psi_1$–summation formula may be formulated not only for $|q| < 1$ but for $|q| = 1$ also.

Diese Ergebnisse lassen vermuten, dass ein bilaterales q-deformiertes Binomialtheorem sowie Ramanujans ${}_1\psi_1$–Summation nicht nur für $|q| < 1$, sondern auch für $|q| = 1$ formuliert werden können.

## 6. Literaturangaben